# Enhancing Near-Field Radiative Heat Transfer between Dissimilar Dielectric Media by Coupling Surface Phonon Polaritons to Graphene's Plasmons

Mehran Habibzadeh[a], Md. Shofiqul Islam[a], Philippe K. Chow[b], and Sheila Edalatpour[a,c*]

[a]Department of Mechanical Engineering, University of Maine, Orono, Maine 04469, USA

[b]Columbia Nano Initiative, Columbia University, New York, New York 10027, USA

[c]Frontier Institute for Research in Sensor Technologies, University of Maine, Orono, Maine 04469, USA

*Email: sheila.edalatpour@maine.edu

## Abstract

Dielectric media are very promising for near-field radiative heat transfer (NFRHT) applications as these materials can thermally emit surface phonon polaritons (SPhPs) resulting in large and quasi-monochromatic heat fluxes. Near-field radiative heat flux between dissimilar dielectric media is much smaller than that between similar dielectric media and is also not quasi-monochromatic. This is due to the mismatch of the SPhP frequencies of the two heat-exchanging dielectric media. Here, we experimentally demonstrate that NFRHT between dissimilar dielectric media increases substantially when a graphene sheet is deposited on the medium with the smaller SPhP frequency. An enhancement of ~ 2.7 to 3.2 folds is measured for the heat flux between SiC and LiF separated by a vacuum gap of size ~ 100 − 140 nm when LiF is covered by a graphene sheet. This enhancement is due to the coupling of SPhPs and surface plasmon polaritons (SPPs). The SPPs of graphene are coupled to the SPhPs of LiF resulting in coupled SPhP-SPPs with a dispersion branch monotonically increasing with the wavevector. This monotonically increasing branch of dispersion relation intersects the dispersion branch of the SPhPs of SiC causing the coupling of the surface modes across the vacuum gap, which resonantly increases the heat flux at the SPhP frequency of SiC. This surface phonon-plasmon coupling also makes NFRHT quasi-monochromatic, which is highly desired for applications such as near-field thermophotovoltaics and thermophotonics. This study experimentally demonstrates that graphene is a very promising material for tuning the magnitude and spectrum of NFRHT between dissimilar dielectric media.

**Keywords:** Near field radiative heat transfer, surface phonon and plasmon coupling, dissimilar dielectrics, graphene, surface phonon polaritons, surface plasmon polaritons

**Introduction**

Radiative heat transfer in the near-field regime (i.e., when the separation gap of the heat-exchanging media is smaller than or comparable to the thermal wavelength) can exceed the far-field blackbody limit by orders of magnitude and be quasi-monochromatic (1,2). The quasi-monochromatic near-field radiative heat transfer (NFRHT) is achieved when the heat-exchanging media thermally emit surface modes such as surface phonon polaritons (SPhPs) and surface plasmon polaritons (SPPs). The enhanced and quasi-monochromatic radiative heat flux in the near-field regime has been capitalized on for several potential applications such as near-field thermophotovoltaic energy harvesting and conversion (e.g., 3), thermal rectification (e.g., 4,5), and near-field photonic cooling (6). Dielectrics are highly promising for NFRHT applications, as these materials can thermally emit SPhPs, which resonantly enhance the heat flux at a given frequency. However, the enhanced and quasi-monochromatic heat flux is achieved between only similar dielectric media. In this case, the dispersion relations of the SPhPs of the two media perfectly match, resulting in a strong coupling between the SPhPs across the vacuum gap. In the case of dissimilar dielectric media, there is a very weak coupling between SPhPs of the two heat-exchanging media resulting in subsided and non-monochromatic heat transfer. It has been theoretically proposed that placing a graphene sheet on one the dielectric media can significantly increase the heat flux due to the interplay between the SPPs of graphene and the SPhPs of the dielectric media (7). However, the enhancement of near-field radiative heat flux between dissimilar dielectric media using graphene has not been experimentally demonstrated yet. The only experimental attempt is concerned with measuring the deflection of an atomic force microscope probe carrying a silica microsphere when the probe approaches a silicon carbide (SiC) plate covered with epitaxial graphene (8). This study is concerned with a microscopic geometry and does not measure the near-field heat flux. The near-field heat flux has been experimentally measured for macroscopic planar media (9-28). However, the enhancement of near-field radiative heat flux between dissimilar dielectrics by utilizing graphene's plasmons has not been experimentally demonstrated yet. In this paper, we experimentally show that the near-field heat flux between macroscopic (surface area of 225 mm$^2$) SiC and lithium fluoride (LiF) plates separated by a nanoscale separation gap of size ~ 100 – 140 nm increases by ~ 2.7 to 3.2 times when the LiF is covered with a graphene sheet. We demonstrate that, unlike the dispersion relation of the SPhPs of bare LiF which has a horizontal asymptote, the dispersion relation of the coupled SPhP-SPPs of the graphene-covered LiF inherits a monotonic increase with the wavevector, $k_\rho$, from graphene's plasmons. As such, the dispersion relation of the coupled SPhP-SPPs can reach and intersect the dispersion relation of the SPhPs of SiC located at a larger frequency. This coupling process results in electromagnetic modes with a large transmission probability and relatively large $k_\rho$s which resonantly increase the heat flux at the SPhP frequency of SiC. The fluctuational electrodynamic simulations of the spectral heat flux show that while the heat flux for the LiF-SiC system has several peaks with the same order of magnitude, the heat flux between the graphene covered LiF and SiC is quasi-monochromatic at the SPhP frequency of SiC. The enhanced and quasi-monochromatic heat flux achieved by utilizing graphene is very promising for future energy conversion and conservation techniques such as near-field thermophotovoltaic (29) and thermophotonic (30) systems as well as thermal management applications such as thermal diodes and rectifiers (31).

**Measuring Radiative Heat Flux Between Two Planar Media**

*Experimental Setup*

A schematic of the experimental setup implemented for measuring the near-field radiative heat flux between two planar media separated by a vacuum gap is shown in Fig. 1a. One of the media, referred to as the emitter hereafter, is heated up using a ceramic heater (HT24S, Thorlabs) which is connected to a power supply (KPS3010D, Eventek). To ensure uniform heating of the emitter, a 20 mm by 20 mm copper plate (grade 110, Grainger) with a thickness of 4.76 mm is placed between the heater and the emitter. To estimate the temperature of the emitter, a hole with a diameter of ~1.5 mm and a depth of ~9.0 mm is drilled laterally into the copper plate and a T-type thermocouple is inserted inside the hole. The emitter is maintained at a distance of $D$ from the second medium using two paperboard posts with a height of 1 mm for the far-field measurements and 361 SU-8 posts with a height varying between 100 and 140 nm for the near-field measurements. The second medium, referred to as the receiver hereafter, is cooled down using a thermoelectric cooler (TEC1-12706, Hebei I. T.). A heat flux meter (PHFS-01e, FluxTeq) is placed beneath the receiver. The heat flux meter is calibrated by the manufacturer using an in-house conduction-based calibration system. Based on the certification of calibration provided by the manufacturer, the heat flux meter provides results which are within 5% of those measured using the conduction-based system. A copper plate (grade 110, Grainger) is inserted between the heat flux meter and the thermoelectric cooler (TEC) to ensures a uniform heat flux from the meter to the TEC. In the same way as for the emitter, a T-type thermocouple is embedded in the copper plate beneath the receiver to monitor the temperature. The T-Type thermocouples are calibrated using a pre-calibrated reference thermocouple, and they are estimated to have an error margin of ±1°C (32). The hot side of the TEC is placed on an aluminum heat sink to dissipate the excess heat into the body of the vacuum chamber, where the experiments take place. Thermal grease (Ceramique 2, Arctic Silver) is applied to all interfaces to reduce the interfacial thermal resistance. The stack is placed inside a U-block, which is 3D printed using Acrylonitrile Butadiene Styrene (ABS) filament. A 2-mm-thick layer of cork is placed on the heater to minimizes conduction from the heater to the U-block. A spring, adjustable through screws on the U-block, is positioned between the cork layer and the U-block. The applied force from the spring to the cork layer keeps the stack in place. A control system was built to maintain the emitter and receiver temperatures at preset values during the experiments by adjusting the current supplied to the heater and the TEC. The setup is assembled in a cleanroom environment and placed in a custom-made vacuum chamber (Kurt J. Lesker Company) pumped down to a pressure between 1 and $9\times10^{-6}$ Torr using a vacuum pump (Turbo-V 301, Varian).

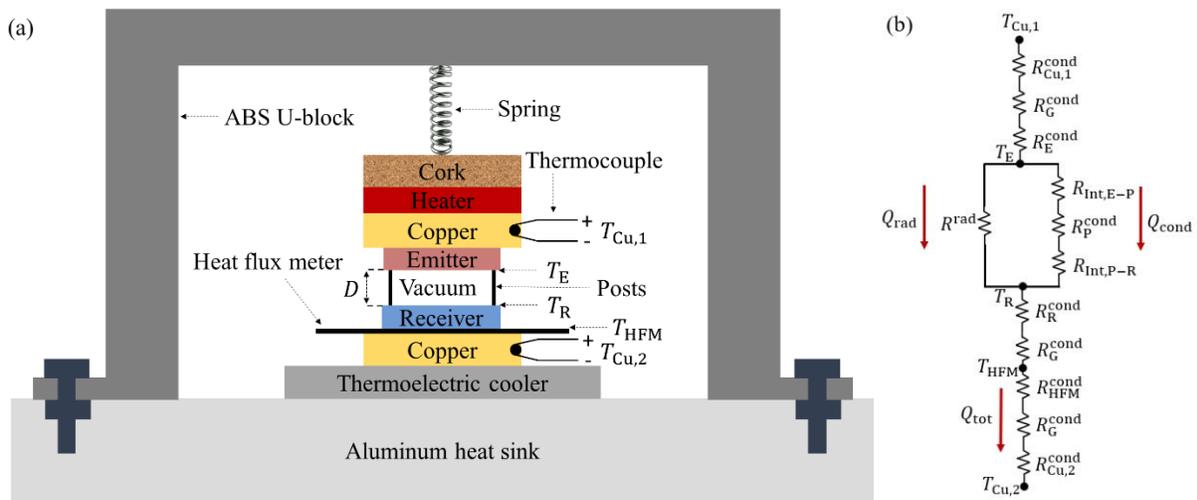

Figure 1 – (a) A schematic of the experimental setup implemented for measuring the near-field radiative heat flux between two planar media. (b) The thermal circuit between the copper heat spreaders on the emitter and receiver sides.

The temperatures of the copper plates on the emitter and receiver sides are set on the desired values. The heater and the TEC are turned on and it is waited until the system reaches a steady state. Then, the temperatures of the copper plates, $T_{\text{Cu},1}$ and $T_{\text{Cu},2}$, and the heat flux, $q_{\text{HFM}}$, are recorded. The thermal circuit of the system is shown in Fig. 1b. The total heat transfer rate through the system can be found as $Q_{\text{tot}} = q_{\text{HFM}}A_{\text{R}}$, where $A_{\text{R}}$ is the surface area of the receiver. The total heat rate, $Q_{\text{tot}}$, is due to the radiative heat rate between the emitter and the receiver, $Q_{\text{rad}}$, as well as conductive heat rate, $Q_{\text{cond}}$, through the spacer posts. The radiative heat rate can be extracted by subtracting the estimated conductive heat rate from the measured total heat rate, i.e.,

$$Q_{\text{rad}} = Q_{\text{tot}} - Q_{\text{cond}} \tag{1}$$

The conductive heat rate through the spacer posts can be estimated as:

$$Q_{\text{cond}} = \frac{T_{\text{E}} - T_{\text{R}}}{R_{\text{Int,E-P}} + R_{\text{P}}^{\text{cond}} + R_{\text{Int,P-R}}} \tag{2}$$

where $R_{\text{Int,E-P}}$, $R_{\text{P}}^{\text{cond}}$, and $R_{\text{Int,P-R}}$ are the thermal resistances of the emitter-posts interface, posts, and posts-receiver interface, respectively, $T_{\text{E}}$ is the temperature of the emitter, and $T_{\text{R}}$ is the temperature of the receiver. As seen from the thermal circuit in Fig. 1b, the temperature $T_{\text{R}}$ can be found from the measured temperature for the heat flux meter as:

$$T_{\text{R}} = T_{\text{HFM}} + Q_{\text{tot}}(R_{\text{R}}^{\text{cond}} + R_{\text{G}}^{\text{cond}}) \tag{3}$$

where $T_{\text{HFM}}$ is the temperature of the top surface of the heat flux meter measured using a T-type thermocouple integrated in the heat flux meter, $R_{\text{R}}^{\text{cond}}$ is the conductive thermal resistance of the receiver, and $R_{\text{G}}^{\text{cond}}$ is the thermal resistance of the grease applied between the receiver and the heat flux meter. The thermal resistance of the receiver is $R_{\text{R}}^{\text{cond}} = \frac{t_{\text{R}}}{k_{\text{R}} A_{\text{R}}}$, where $t_{\text{R}}$ and $k_{\text{R}}$ are the thickness and thermal conductivity of the receiver, respectively. The thermal resistance of the grease layers can be estimated by using the measured heat flux $q_{\text{HFM}}$ and the temperatures of the copper plate on the receiver side, $T_{\text{Cu,2}}$, and heat flux meter, $T_{\text{HFM}}$, as:

$$R_{\text{G}}^{\text{cond}} = \frac{T_{\text{HFM}} - T_{\text{Cu,2}}}{Q_{\text{tot}}} - \left(R_{\text{HFM}}^{\text{cond}} + R_{\text{Cu,2}}^{\text{cond}}\right) \qquad (4)$$

where $R_{\text{HFM}}^{\text{cond}} = 1$ K/W is the thermal resistance of the heat flux meter as reported by the manufacture. The thermal resistance of the copper plate is $R_{\text{Cu,2}}^{\text{cond}} = \frac{0.5 t_{\text{Cu,2}}}{k_{\text{Cu,2}} A_{\text{Cu,2}}}$ where $t_{\text{Cu,2}}$, $k_{\text{Cu,2}}$, and $A_{\text{Cu,2}}$ are the thickness (= 4.76 mm), thermal conductivity (= 387 W/mK (33)), and surface area (= 400 mm$^2$) of the copper plate, respectively, and the factor 0.5 accounts for the fact that the thermocouple is located in the middle of the copper plate.

The temperature of the emitter $T_{\text{E}}$ in Eq. 2 can be found from the measured $T_{\text{Cu,1}}$ as:

$$T_{\text{E}} = T_{\text{Cu,1}} - Q_{\text{tot}}(R_{\text{Cu,1}}^{\text{cond}} + R_{\text{G}}^{\text{cond}} + R_{\text{E}}^{\text{cond}}) \qquad (5)$$

where $R_{\text{Cu,1}}^{\text{cond}}$ is the thermal resistance of the copper plate on the emitter side, $R_{\text{G}}^{\text{cond}}$ is the thermal resistance of the grease applied between the copper plate and the emitter, and $R_{\text{E}}^{\text{cond}}$ is the conductive thermal resistance of the emitter. The thermal resistance of the copper plate $R_{\text{Cu,1}}^{\text{cond}}$ can be found in the same way as $R_{\text{Cu,2}}^{\text{cond}}$. The thermal resistance of the emitter is $R_{\text{E}}^{\text{cond}} = \frac{t_{\text{E}}}{k_{\text{E}} A_{\text{E}}}$, where $t_{\text{E}}$, $k_{\text{E}}$, and $A_{\text{E}}$ are the thickness, thermal conductivity, and surface area of the emitter, respectively.

**Results**

*Far-Field Radiative Heat Transfer between Two Blackbodies*

In this sub-section, we use the implemented experimental setup to measure the radiative heat flux between two planar blackbodies separated by a vacuum gap of size 1 mm. We compare the measured heat fluxes with the theoretical predictions. For this purpose, two copper plates, each having a surface area of 20 by 20 mm$^2$ and a thickness of 4.76 mm, are painted with a blackbody paint (SP102, VHT). The temperatures of the thermometers embedded in the copper plates are set to $T_{\text{Cu,1}} = 50°C$ and $T_{\text{Cu,2}} = 20°C$. Two paperboard posts with a low thermal conductivity of $k_P = 0.12$ W/mK [34], a total cross-sectional area of $A_P = 12$ mm², and a thickness of $t_P = 1$ mm are placed between the two blackbodies to create a 1 mm gap between the two media. The vacuum chamber is pumped down to a pressure of $6.4 \times 10^{-6}$ Torr. Heat is transferred from the hot blackbody to the cold one via radiation as well as conduction through the paperboard posts. The heat flux has been measured twice. For each measurement, the heater and the TEC are turned on and it is waited until the system reaches a steady state at which point the heat flux meter is read. Then, the heater and the TEC are turned off, and it is waited until the setup reaches the ambient temperature. After thermal equilibrium is achieved, the second round of measurements is taken in the same way as the first one. The recorded heat fluxes are $q_{\text{HFM}} = 219.2$ and 219.5 W/m², which are shown versus the difference between the measured $T_{\text{Cu,1}}$ and $T_{\text{HFM}}$ in Fig. 2a.

The radiative portion of the measured heat flux, $q_{\text{rad}} = \frac{Q_{\text{rad}}}{A_E}$, is found using Eq. 1 and by estimating the heat conduction through the paperboard posts, $Q_{\text{cond}}$, using Eq. 2. The thermal resistance of the grease is found using Eq. 4 as $R_G^{\text{cond}} = 26.35$ and 26.43 K/W for the first and second experiment, respectively. The thermal resistances of the emitter and receiver blackbodies are found using the thermal conductivity, surface area and thickness of the copper plate as reported before. The temperatures of the blackbodies are found as $T_E = 47.56$ K and $T_R = 24.81$ K for the first measurement and $T_E = 47.69$ K and $T_R = 24.62$ K for the second measurement. The interfacial thermal resistances between the blackbodies and the paperboard posts, $R_{\text{Int,E−P}}$ and $R_{\text{Int,P−R}}$, which is less than 0.8 K/W (35), are negligible compared to the thermal resistance of the posts ($R_P^{\text{cond}} = 694.4$ K/W) and thus are neglected. The conductive heat transfer through the paperboard posts is found using Eq. 2 as $Q_{cond}$ = 0.0327 and 0.0331 W for the first and second measurements, respectively. Using the estimated $Q_{cond}$, the radiative heat flux between blackbodies, $q_{rad}$, is found using Eq. 1 as 141.5 and 140.7 W/m$^2$ for the first and second measurements, respectively. The measured radiative heat flux versus the temperature difference of the emitter and receiver, $T_E - T_R$, is shown in Fig. 2b. To compare measurements against the theory, the far-field radiative heat flux between the two blackbodies is modeled as $q_{rad} = \frac{\sigma(T_E^4 - T_R^4)}{\frac{1-\varepsilon_E}{\varepsilon_E} + \frac{1}{F_{E-R}} + \frac{1-\varepsilon_R}{\varepsilon_R}}$ (36). In this equation, $\sigma$ = 5.67×10$^{-8}$ Wm$^{-2}$K$^{-4}$ is the Stefan–Boltzmann constant, $\varepsilon_E$ and $\varepsilon_R$ are the emissivities of the blackbodies, and $F_{E-R}$ = 0.9079 is the view factor between the two blackbodies found using the equations presented in Ref. (37). Considering an emissivity of 0.99 for the blackbodies, $q_{rad}$ is calculated for a temperature difference, $\Delta T = T_E - T_R$, range of 5 to 60°C and is compared with the experimental data points in Fig. 2b. The predicted radiative heat fluxes for $\Delta T$ = 23.07 and 22.75°C are $q_{rad}$ = 136.1 and 138.1 W/m$^2$, respectively, which are different from the experimentally measured values by 4.0 and 1.9 %, respectively. This small difference between the theoretical and experimental data shows the capability of the experimental setup for estimating the radiative heat transfer between two planar media.

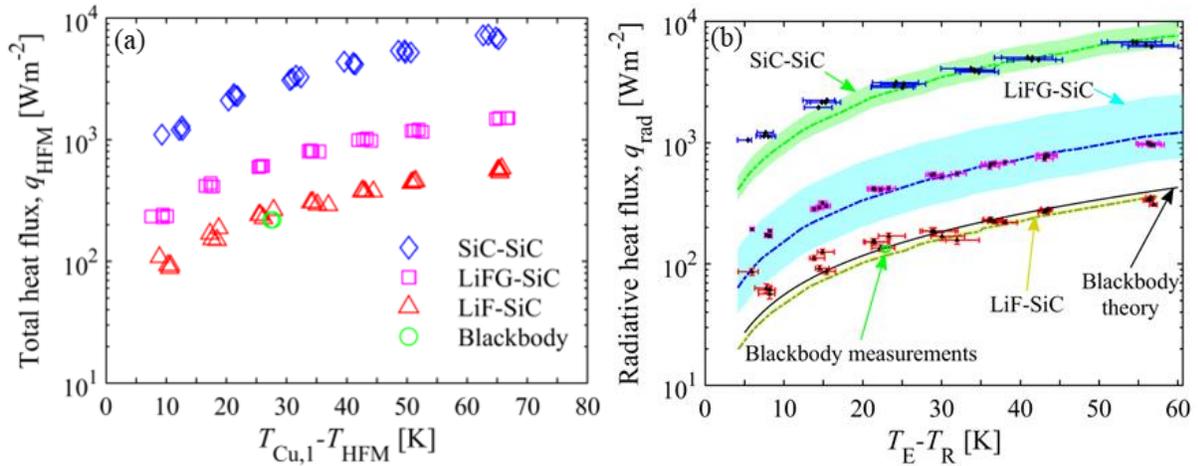

Figure 2 – (a) The total (radiative and conductive) heat flux measured by the heat flux meter, $q_{\text{HFM}}$, versus the difference between the measured temperatures for the copper plate at the emitter side and the heat flux meter, $T_{\text{Cu,1}} - T_{\text{HFM}}$. (b) The radiative portion of the measured heat flux, $q_{\text{rad}}$, versus the temperature difference between the emitter and the receiver, $T_{\text{E}} - T_{\text{R}}$. The symbols show the experimental measurements while the colored bands represent the theoretically predicted heat flux for a gap size range of $D$, in the range of $100 - 140$ nm. The dash lines show the theoretical heat flux for an average gap size of $D = 120$ nm.

*Near-Field Radiative Heat Transfer between Two SiC plates*

In this sub-section, the experimental setup is utilized for measuring the near-field radiative heat transfer between two similar dielectric media, namely two SiC plates, separated by a nanoscale vacuum gap. The measurements are compared against theoretical predictions using the fluctuational electrodynamics. The SiC plates have a surface area of 15 mm by 15 mm and a thickness of 0.43 mm. A total of 361 SU-8 posts with a height varying between $100 - 140$ nm and a cross-sectional area of 9 μm² are fabricated on one of the SiC plates to maintain a nanoscale gap between the hot and cold sides (see Methods for the fabrication process). Six temperature differences across the vacuum gap, $\Delta T$, ranging from ~ 5 to 60 K are considered for the experiments. The experiment is repeated four times for each considered temperature difference resulting in a total of 24 data points. The pressure of the vacuum chamber for these measurements is kept between 1 to $9 \times 10^{-6}$ Torr. The first round of the measurements is done consecutively for all considered temperature differences starting from the smallest to the largest. Then, the heater and the TEC are turned off, and the set-up is allowed to reach thermal equilibrium with the environment. Next, the heater and the TEC are turned on and the second round of the measurements is taken for all considered temperature differences. After the second round of measurements, the setup is allowed to cool down. Then, the set-up removed from the vacuum chamber and is completely disassembled. The setup is reassembled and the third and fourth rounds of the measurements are conducted in the same manner as the first and second ones. The measured heat fluxes are shown versus the difference between the measured $T_{\mathrm{Cu,1}}$ and $T_{\mathrm{HFM}}$ in Fig. 2a.

The near-field radiative heat flux, $q_{\mathrm{rad}}$, is found from the measured heat flux, $q_{\mathrm{HFM}}$, using Eqs. 1-5. When predicting the thermal resistances of the receiver and emitter, a thermal conductivity of $k_{\mathrm{SiC}} = 320$ W/mK is considered for SiC [38]. A thermal conductivity of $k_{\mathrm{P}} = 0.2$ W/mK [39] and an average thickness of 120 nm are assumed for the SU-8 posts when estimating their thermal resistance. Since the SU-8 posts are fabricated on the receiver, $R_{\mathrm{Int,P-R}} \approx 0$ is assumed [20]. The contact thermal resistance at the emitter-post interface, $R_{\mathrm{Int,E-P}}$, depends on the thermal conductivities of the emitter and posts as well as the smoothness of the surfaces and the applied pressure. As such, $R_{\mathrm{Int,E-P}}$ is considered as a fitting parameter [12]. The fitted value for $R_{\mathrm{Int,E-P}}$ is $1 \times 10^{-6}$ m$^2$K/W, which is within the range reported in the literature [35]. For each of the six considered temperature differences, four thermal resistances are found for the grease layer at the copper-heat flux meter interface (using Eq. 4) corresponding to the four repetitions of the experiment. The thermal resistance of the grease at the copper-emitter and receiver-heat flux meter interfaces are then assumed to vary within the range spanned by these four estimated values. The near-field radiative heat flux, found using the measured heat flux and Eq. 1, is shown versus the temperature difference in Fig. 2b. The uncertainty associated with the thermal resistance of the grease, which affects the estimation of $T_{\mathrm{R}}$ and $T_{\mathrm{E}}$ as found using Eqs. 3 and 4, is shown by error bars in Fig. 2b. The theoretical near-field heat flux predicted using fluctuational electrodynamics for gap sizes ranging from $D = 100$ nm to 140 nm are also shown in Fig. 2b (see Methods for details of the theoretical model). The dielectric function of SiC is modeled using the Lorentz oscillator as $\varepsilon_r(\omega) = \varepsilon_{r,\infty}\frac{\omega^2 - \omega_{\text{LO}}^2 + i\Gamma\omega}{\omega^2 - \omega_{\text{TO}}^2 + i\Gamma\omega}$, where $\varepsilon_{r,\infty} = 6.46$, $\omega_{\text{LO}} = 18.30\times10^{13}$ rad/s, and $\omega_{\text{TO}} = 15.01\times10^{13}$ rad/s (40). A phonon scattering rate of $\Gamma = 8.97\times10^{11}$ rad/s is assumed for the SiC samples (41). Figure 2b shows that the near-field radiative heat flux between SiC plates exceeds the blackbody limit by $16.7 - 26.5$ times (depending on the temperature difference, $\Delta T$), which, as will be explained later, is due to the strong coupling of SPhPs thermally excited at the SiC-vacuum interfaces.

<u>*Near-Field Radiative Heat Transfer between LiF and SiC*</u>

In this sub-section, we utilize the implemented experimental setup to demonstrate that the near-field radiative transfer between two dissimilar dielectric media, namely SiC and LiF, can be enhanced by placing a graphene sheet on the LiF substrate, which supports the SPhPs at a lower frequency than SiC. SiC and LiF are selected for this study for a few reasons. First, both materials support SPhPs in the mid-infrared, where these modes can be thermally excited at low to moderate temperatures. Additionally, the coupling of SPhPs of these two materials with SPPs of graphene, and thus heat flux enhancement, can be achieved with small chemical potentials of graphene thus eliminating the need for gating or doping graphene. Finally, research grade SiC and LiF can be purchased via commercial suppliers. The SiC sample on which the SU-8 posts are fabricated serves as the receiver for this experiment, while the LiF plate is considered as the emitter. Two LiF samples (Stanford Advanced Materials and Biotain Crystal) were used for the measurements. Both LiF samples have the same surface area of 15 mm by 15 mm as the SiC sample. One of the LiF samples (Stanford Advanced Materials) is 0.53-mm thick, while the other (Biotain Crystal) has a thickness of 0.5 mm. The heat flux between the two samples is measured in the absence and in the presence of a monolayer graphene sheet (Product#: ME0613, MSE supply) for seven temperature differences ranging from $\Delta T \approx 5$ to 60 K. The graphene sheet is transferred onto the LiF substrate using the standard wet transfer method (See Methods for the details of the transfer technique). Similar to the SiC-SiC case, the heat flux is measured four times for each of the seven considered temperature differences. After the first two rounds of the measurements, the setup is removed from the vacuum chamber and disassembled. Then, a new graphene layer is transferred onto a new LiF sample for the third and fourth rounds of the measurements to ensure the proper transfer process as well as the reproducibility of the measurements. The measured heat fluxes are shown versus the difference between the measured $T_{\mathrm{Cu,1}}$ and $T_{\mathrm{HFM}}$ in Fig. 2a.

The near-field heat flux is found from the measured heat flux data using Eqs. 1-5. When using Eqs. 1-5, a thermal conductivity of $k_{\mathrm{LiF}}$ = 13.89 W/mK (42) is assumed for the LiF sample, and an interfacial thermal resistance of $R_{\mathrm{Int,E-P}} = 3\times10^{-6}$ m$^2$K/W is found at the interface of LiF and SU-8 posts. Graphene reduces the interfacial thermal resistance (43,44). As such, $R_{\mathrm{Int,E-P}} = 1\times10^{-6}$ m$^2$K/W is assumed for the case when graphene is present. The measured near-field radiative heat flux in the absence and presence of graphene is shown in Fig. 2b. The horizontal error bars are associated with the uncertainty in the thermal resistance of grease, which is predicted in the same manner as explained for the SiC-SiC case. The near-field heat flux is also theoretically predicted for a gap size range of 100 – 140 nm in the presence and absence of graphene (see Methods for details) and is presented in Fig. 2b. The Lorentz oscillator model is used for the dielectric function of LiF. In the Lorenz model, $\varepsilon_{r,\infty}$ = 1.90, $\omega_{\mathrm{LO}} = 12\times10^{13}$ rad/s, $\omega_{\mathrm{TO}} = 5.83\times10^{13}$ rad/s, and $\Gamma = 8.97\times10^{11}$ rad/s are assumed (45). When graphene is hosted by polar dielectric materials, the Fermi level of the graphene can shift away from the Dirac point due to the charged impurities in the dielectric substrate (46). For this reason, for the theoretical predictions of near-field heat flux in the presence of graphene, the chemical potential of graphene, $\mu_c$, is considered as the fitting parameter and is found as 0.17 eV. The scattering rate of the electrons in graphene, $\gamma$, can then be found using the chemical potential as $\gamma = \frac{q_e v_F^2}{\mu_e \mu_c}$ (47,48), where $v_F = 9.5 \times 10^5$ m/s is the Fermi velocity (47,48) and a carrier mobility of $\mu_e = 5000$ cm$^2$/Vs is assumed based on the manufacture data. The potential sources of difference between theory and experiment are the uncertainty in the exact values of the thermal conductivity and height of the SU-8 posts, the thermal conductivity of the grease, and the uniformity of the temperature of the surfaces. These errors can be reduced or mitigated by using spacer posts with known thermal conductivity (24), a more advanced vacuum gap system such as a nano-positioner (19), thermal pads instead of thermal grease (21), and an average temperature taken over the surface, respectively. Nevertheless, it is seen from Fig. 2b that overall, the measurements are in good agreement with the theoretical predictions.

**Discussion**

Figure 2b shows that the near-field radiative heat flux in the case of dissimilar dielectrics, i.e., for the LiF-SiC system, is much smaller than that between two similar SiC plates. The near-field heat flux for the LiF-SiC system is even smaller than the far-field heat flux between two blackbodies. The near-field radiative heat flux between LiF and SiC increases by 2.7 to 3.2 times (depending on the temperature difference), exceeding the blackbody limit, when LiF is covered with a graphene sheet. To understand the physics underlying the enhancement of heat flux in the presence of the graphene, the spectral heat flux, $q_{\mathrm{rad},\omega}$, and the spectral heat flux per unit $k_\rho$, $q_{\mathrm{rad},\omega,k_\rho}$, are modeled for two LiF plates, two SiC plates, and a LiF and a SiC plate. The emitter is assumed at $T_{\mathrm{E}} = 328.5$ K, while a temperature of $T_{\mathrm{R}} = 296.6$ K is assumed for the receiver. The heat flux is computed for the average gap size of $D = 120$ nm. Figure 3a compares $q_{\mathrm{rad},\omega}$ for the three cases, while $q_{\mathrm{rad},\omega,k_\rho}$ is presented in Figs. 3b-d for the LiF-LiF, SiC-SiC, and LiF-SiC systems, respectively. It should be noted that $q_{\text{rad},\omega,k_\rho}$ plotted in Figs. 3b-d only includes the contribution from the transverse magnetic (TM) polarization, as the heat flux in all three cases is driven by the TM-polarized electromagnetic waves (2,49). The dispersion relations of the SPhPs are also plotted in Figs. 3b-d (see Methods for details). It is seen from Fig. 3a that the heat flux for LiF-LiF and SiC-SiC cases, where emitter and receiver are made from similar materials, is quasi-monochromatic and is completely dominated by the contribution from a small spectral band around $\omega_{\text{SPhP, LiF}} = 1.03{\times}10^{14}$ rad/s and $\omega_{\text{SPhP, SiC}} = 1.79{\times}10^{14}$ rad/s, respectively. The origin of these peaks, which is well understood, is thermal excitation of SPhPs supported by the LiF and SiC plates. The dispersion relations of the SPhPs for single interfaces of LiF and SiC (i.e., for vacuum-LiF and vacuum-SiC interfaces) as well as for the LiF-LiF and SiC-SiC systems are plotted in Fig. 3b and 3c. For LiF-LiF and SiC-SiC systems, the dispersion relations of the SPhPs of the emitter and receiver overlap resulting in a strong coupling between the SPhPs of the two media. Due to this strong coupling, the dispersion relations of the SPhPs split into a symmetric and an antisymmetric branch, which converge to the asymptote of the dispersion relation for a single interface (i.e., to $\omega = \omega_{\text{SPhP}}$) at large wavevectors. Due to the very strong contribution of SPhPs to the heat flux at $\omega_{\text{SPhP}}$, the heat flux is monochromatically enhanced at this resonance frequency. It should also be mentioned that the heat flux spectral for both SiC-SiC and LiF-LiF systems have an additional peak around $\omega_{\text{TO}}$. This peak, which has a magnitude much smaller than the SPhP peak, is due to a local maximum in the imaginary part of the dielectric function, Im[$\varepsilon_r$], at this frequency. Based on fluctuational electrodynamics, the spatial correlation of the thermally-generated stochastic current density is directly proportional to Im[$\varepsilon_r$] (50). The spectral heat flux for the LiF-SiC case, where the emitter and receiver are not made from similar materials, is also shown in Fig. 3a. It is seen that the heat flux between these dissimilar dielectric media is significantly (more than one orders of magnitude) smaller than that between the similar dielectric media. Additionally, the heat flux between LiF and SiC plates is not quasi-monochromatic and has four peaks with comparable magnitudes at $5.56 \times 10^{13}$, $1.02 \times 10^{14}$, $1.46 \times 10^{14}$, and $1.79 \times 10^{14}$ rad/s corresponding to $\omega_{TO,LiF}$, $\omega_{SPhP,LiF}$, $\omega_{TO,SiC}$, and $\omega_{SPhP,SiC}$, respectively. The reason for the small and non-monochromatic heat flux can be explained by considering the dispersion relations of SPhPs for the LiF-SiC system as shown in Fig. 3d. It is seen from Fig. 3d that there is only a weak coupling between the SPhPs of SiC and LiF which occurs at small frequencies ($\omega < 1 \times 10^{14}$ rad/s) and wavevectors ($k_\rho < k_0$). The LiF branch of dispersion relation for the LiF-SiC system is only slightly different from the dispersion relation for a single interface of LiF. Both LiF and SiC branches of dispersion relation for the LiF-SiC system maintain their horizonal asymptote at $\omega = \omega_{SPhP,LiF}$ and $\omega_{SPhP,SiC}$, respectively. Due to the weak coupling between the SPhPs of the LiF and SiC, the heat flux for the LiF-SiC case is significantly lower than that for LiF-LiF and SiC-SiC cases and is not monochromatic.

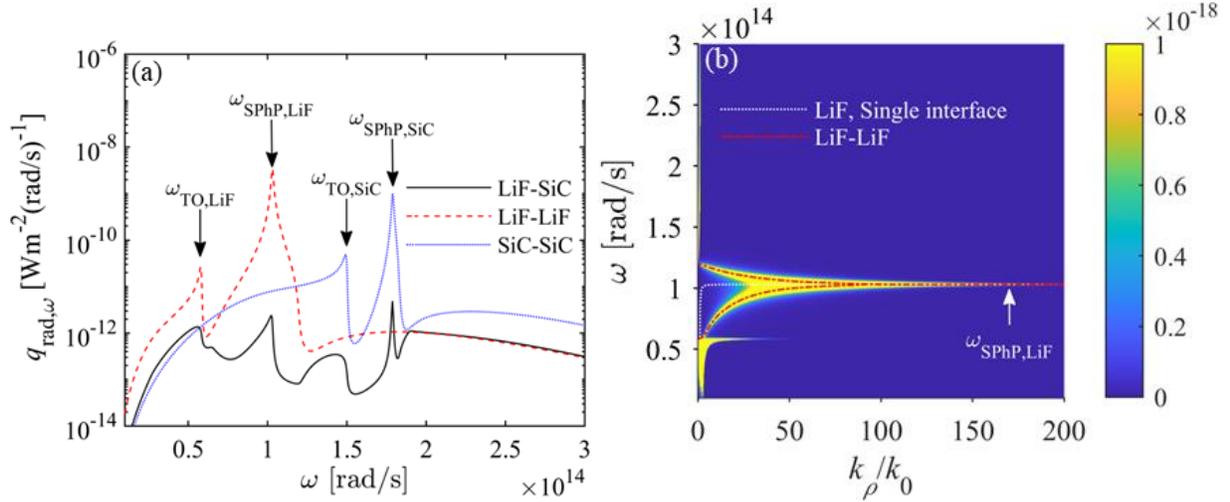

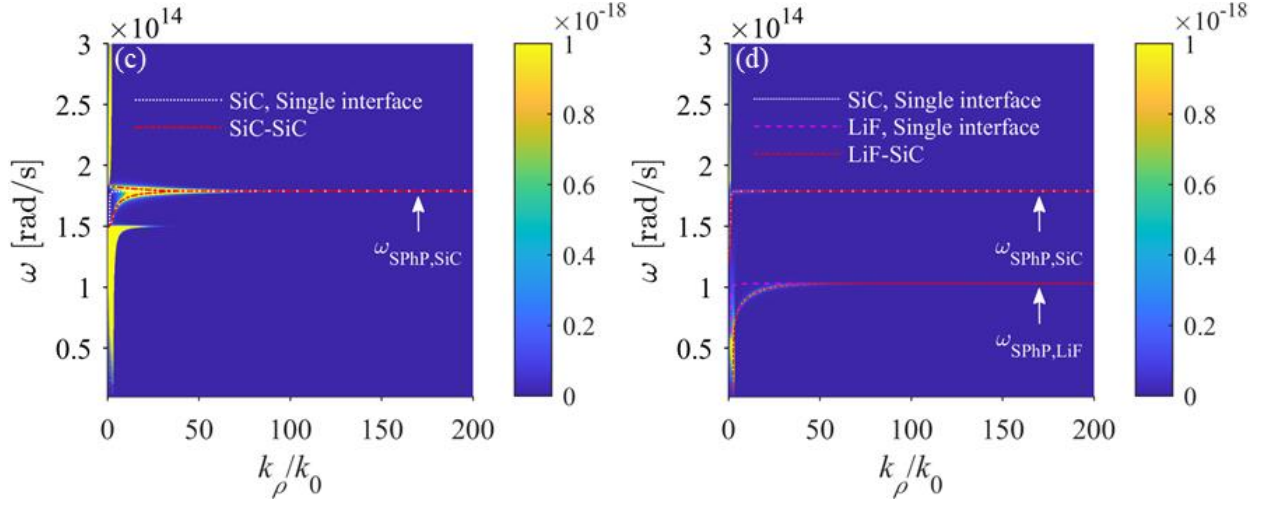

Figure 3 – (a) The spectral heat flux, $q_{rad,\omega}$, theoretically predicted for LiF-LiF, SiC-SiC, and LiF-SiC systems. (b) The spectral heat flux per unit $k_\rho$ mediated by the TM-polarized electromagnetic waves, $q_{rad,\omega,k_\rho}$, for (b,c,d) LiF-LiF, (c) SiC-SiC, and (d) LiF-SiC systems. The receiver is assumed at 296.6 K while the emitter has a temperature of 328.5 K. An average vacuum gap size of $D =$ 120 nm is assumed. The dispersion relations are also plotted in Panels b-d. The unit for the color bars in Panels b-d is $Wm^{-2}(rad/s)^{-1}m$.

The spectral heat flux in the presence of graphene sheet on LiF (LiFG) is compared with the one in the absence of graphene in Fig. 4a. The spectral heat flux per unit $k_\rho$ for the LiFG-SiC system and the dispersion relations of the SPhPs are shown in Fig. 4b for both LiF-SiC and LiFG-SiC cases. The magnitude of the heat flux at $\omega_{SPhP,SiC}$ increases by ∼ 44 times when LiF is covered with the graphene sheet. There is also a broadband enhancement for the heat flux in the spectral range of ∼1×10$^{14}$ to 1.9×10$^{14}$ rad/s compared to the case with no graphene sheet. However, the heat flux for LiFG-SiC system is completely dominated by the contribution from SPhP modes located around $\omega_{SPhP,SiC}$ and is quasi-monochromatic at this frequency (see the inset of Fig. 4a for $q_{rad,\omega}$ versus $\omega$ in a linear scale). The reason for the enhancement of heat flux in the presence of graphene can be explained using Fig. 4b. It is seen from this figure that the SPhPs of LiF couple to the SPPs of graphene and they split into two branches. The upper branch, unlike the SPhP branch of LiF for the LiF-SiC system, does not have a horizontal asymptote and inherits monotonically increasing behavior from graphene plasmons (7). The dispersion branch associated with the SPhPs of SiC, however, retain its horizontal asymptote at $\omega_{SPhP,SiC}$, since the vacuum gap prevents a strong coupling with the SPPs of graphene. The monotonically increasing dispersion branch intersects the SiC branch of SPhPs at relatively large $k_\rho$s ($\sim 60\ k_0$), resulting in a region of highly contributing modes at $\omega_{SPhP,SiC}$. This coupling process enhances the heat flux at $\omega_{SPhP,SiC}$ by 44 times causing a monochromatic heat flux at this frequency. It is also seen from Fig. 4b that the broadband enhancement from $1 \times 10^{14}$ to $1.9 \times 10^{14}$ rad/s is due to the upper branch of the coupled SPhP-SPPs associated with LiFG. The spectral heat flux in Fig. 4a demonstrates that covering the dielectric medium with the smaller SPhP frequency with a graphene sheet can substantially and quasi-monochromatically increases the heat flux due to the strong coupling of surface plasmon and phonon polaritons.

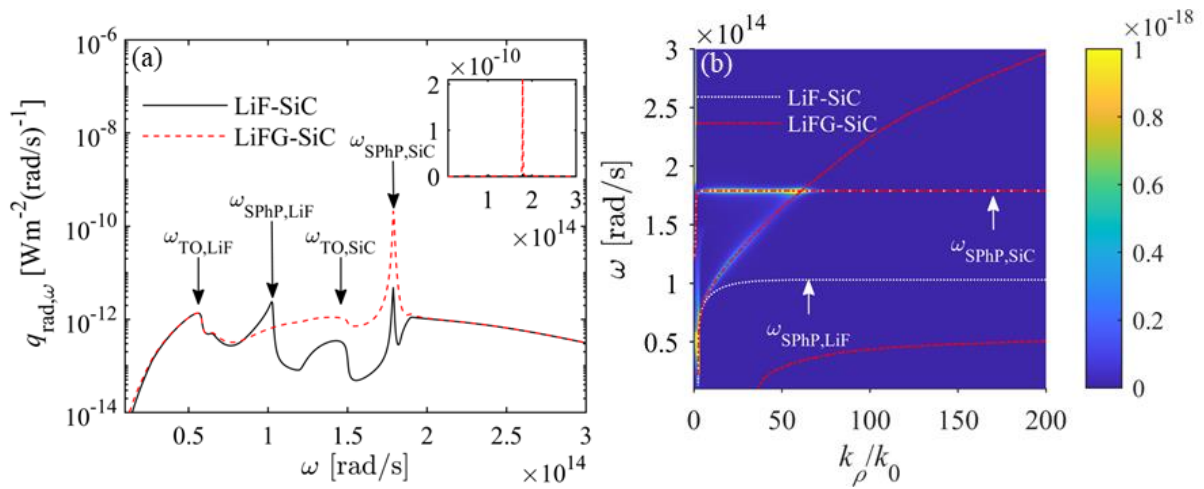

Figure 4 – (a) Spectral radiative heat flux, $q_{\mathrm{rad},\omega}$, theoretically predicted between LiF and SiC plates in comparison with that predicted between graphene-covered LiF (LiFG) and SiC. The inset shows the spectral heat flux versus frequency in a linear scale. (b) The spectral heat flux per unit $k_\rho$, $q_{\mathrm{rad},\omega,k_\rho}$, mediated by the TM-polarized electromagnetic waves for the LiFG-SiC system. The dispersion relation for the LiF-SiC and LiFG-SiC systems are also plotted in Panel b. A receiver (SiC) temperature of 296.6 K, an emitter (LiF and LiFG) temperature of 328.5 K, and an average gap size of $D = 120$ nm are assumed. The unit of the color bar in Panel b is $\mathrm{Wm^{-2}(rad/s)^{-1}m}$.

The enhancement of the heat flux between LiF and SiC in the presence of the graphene sheet, which has a chemical potential of $\mu_c = 0.17$ eV, is measured to be between 2.7-3.2 times depending on the temperature difference. The maximal enhancement of the heat flux for these two materials is between 7.4 and 7.7, depending on the temperature difference, which is achieved for an optimal chemical potential of 0.29 eV. When $\mu_c = 0.29$ eV, the upper dispersion branch of the coupled SPP-SPhPs acquires an optimal slope and intersect the dispersion branch associated with the SPhPs of SiC around the wavevector of the surface modes with the largest contribution to the heat flux, $k_{\rho,\mathrm{max}}$. When the misalignment between the resonances of the emitter and receiver increases, $k_{\rho,\mathrm{max}}$ decreases. As such, a greater slope for the upper branch of the coupled SPhP-SPPs associated with graphene-on-substrate is required to intersect the SPhP dispersion branch of the material across the vacuum gap around $k_{\rho,\mathrm{max}}$. To increase the slope of the upper branch of the coupled SPhP-SPPs, a larger chemical potential is needed (See Section A of the Supporting Information for more details).

While LiF and SiC are selected for this study, the enhanced and quasi-monochromatic heat flux obtained using graphene is not limited to these two dielectric materials. For example, we studied the maximal enhancement factor that can be obtained between LiF and eleven other materials supporting SPhPs at various frequencies (see Section B of Supporting Information). The maximal enhancement factor is plotted in Figure S3a of the Supporting Information. Figure S3a shows that enhancement factor varies between 1.1 and 25.4 depending on the misalignment of the SPhP frequencies of the emitter and receiver. The smaller enhancement factors belong to the cases where the dispersion relation of the SPhPs of the receiver is located between the two branches of the coupled SPhP-SPPs of the graphene-covered LiF (LiFG), and thus cannot intersect any of these two branches (e.g., see the dispersion relation for LiFG-InP in Fig. S3b of the Supporting Information). The greater enhancement factors are obtained for cases where there is a large mismatch between the SPhP resonances of the two materials and the SPhP branch of the receiver intersects one of the two branches of the coupled SPhP-SPPs associated with LiFG at large wavevectors. (e.g., see the dispersion relation for LiFG-GaN and LiFG-BaF$_2$in Figs. S3c and S3d of the Supporting Information, respectively).

While a single layer of graphene is used in this study, larger enhancement factors can be achieved by using multiple layers of graphene. When the number of graphene layers increases, the lower branch of the dispersion relation associated with LiFG also acquires a positive slope (see Fig. S4c in Supporting Information). The positive slope of the lower dispersion branch enables its coupling with the dispersion branch of the SPhPs of SiC, increasing the heat flux at the resonance frequency of SiC (See Fig. S4c in Supporting Information). It should also be mentioned that the enhancement factor eventually saturates when the number of graphene layers increases. As the number of graphene layers increases, the contribution of graphene layers adjacent to the LiF emitter to the heat transfer decreases and eventually vanishes (See Section C of Supporting Information for more details).

## Conclusions

We experimentally measured the near-field radiative heat flux between two macroscopic planar media made of dielectric materials separated by a vacuum gap of size ~100 – 140 nm. The experiments were performed for two SiC plates, a SiC and a LiF plate, and a graphene covered LiF plate and a SiC plate. The measurements showed that the near-field radiative heat flux between dissimilar dielectric media (i.e., between LiF and SiC) is significantly smaller than between similar dielectric media (i.e., between SiC and SiC). This is due to the mismatch of the surface phonon polariton (SPhP) frequencies of SiC and LiF, which does not allow for a strong coupling between the SPhP modes of these two media across the vacuum gap. We experimentally demonstrated that the near-field radiative heat flux between LiF and SiC increases by ~2.7 to 3.2 folds, depending on the temperature difference, when LiF is covered with a graphene sheet. In this case, the surface plasmon polaritons (SPPs) of graphene couple to the SPhPs of LiF, resulting in a branch of coupled SPP-SPhPs which monotonically increases with the wavevector and intersects the SiC branch of the dispersion relation. This coupling process results in highly efficient electromagnetic modes with relatively large wavevectors that increase the heat flux compared to the LiF-SiC case where a very weak coupling between SPhPs exists. It is also seen that the heat flux between LiF and SiC in the presence of graphene is quasi-monochromatic at the SPhP frequency of SiC as opposed to the case when no graphene sheet is used. Our study demonstrated the potential of graphene for achieving enhanced and quasi-monochromatic near-field heat flux between media with mismatching surface resonances, which can benefit thermal management applications and energy conversion technologies such as thermophotovoltaics and thermophotonics.

## Methods

### *Fabricating SU-8 Posts on a SiC Plate*

A nanoscale vacuum gap is maintained between the two planar dielectric media for near-field radiative heat transfer measurements. The gap is generated by fabricating a total of 361 SU-8 posts each with a cross-sectional area of 9 $\mu m^2$ and a nanometer scale height on the surface of the receiver (a 6H-SiC chip). To fabricate the SU-8 posts, 6H-SiC wafers (430-$\mu$m-thick, MSE Supplies) are diced into 15 $mm^2$ chips and cleaned in piranha solution. SU-8 3005 (Kayaku Advanced Materials) is diluted with SU-8 Thinner in a 1:4 ratio, filtered to 0.45 $\mu m$, and spin-coated onto the SiC at 3500 rpm. The chip is soft-baked at 95℃ for 60 seconds. Then, a layer of anti-charging solution (Dischem, DisCharge X2) is spin-coated at 6000 rpm. The 9 $\mu m^2$ features are exposed in a Thermo Fisher Nova NanoSEM 450 equipped with Nabity Pattern Generation System (NPGS) using a 30 kV acceleration voltage, 30 $\mu m$ aperture and 660 pA beam current. The exposure dose is 50 $\mu$C/cm$^2$. The individual posts are patterned in a square array with a pitch of 760 $\mu$m. The 330 $\mu$m edge region of the chip is left unpatterned. After exposure, a post-exposure bake, also at 95℃ for 60 seconds, is performed to reveal the patterns. The sample is then developed in SU-8 developer for 60 seconds, using ultrasonication in the last 10 seconds to reduce residue, rinsed in isopropanol and dried with nitrogen. The heights of several fabricated posts are measured using an atomic force microscope probe. The measured heights for the posts located at the center of the SiC plate vary between 100 nm and 140 nm, while the posts located at the corners and edges have a height between 133 and 208 nm.

*Theoretical Modeling of Near-Field Radiative Heat Flux using Fluctuational Electrodynamics*

The near-field radiative heat flux through the vacuum gap is modeled using fluctuational electrodynamics and by utilizing the scattering matrix approach (47). The schematic considered for theoretical modeling of the system is shown in Fig. 5. The emitter (medium E) with a thickness of $t_E$ is separated from the receiver (medium R) with a thickness of $t_R$ by a vacuum gap (medium

0) of size $D$. The temperatures of the emitter and receiver are $T_E$ and $T_R$, respectively. The copper plate on the emitter side is optically thick, and thus is modeled as a semi-infinite medium (medium s). The receiver is placed on the heat flux meter, which is encapsulated in copper. As such, a semi-infinite copper substate (medium s) is also assumed for the receiver. The temperatures of the semi-infinite copper plates are approximated to be the same as those of the emitter and receiver. The heat flux can be obtained by integrating the spectral (i.e., frequency-dependent) heat flux, $q_{\mathrm{rad},\omega}$, as $q_{\mathrm{rad}} = \int_0^\infty q_{\mathrm{rad},\omega} d\omega$, where $\omega$ is the angular frequency. The spectral heat flux can be found using the transmission functions of propagating and evanescent waves as (47):

$$q_{\mathrm{rad},\omega} = \frac{\Theta(\omega, T_R) - \Theta(\omega, T_E)}{4\pi^2} \sum_{\alpha = \mathrm{TE,\,TM}} \left( \int_0^{k_0} \zeta_{\mathrm{prop}}^\alpha k_\rho dk_\rho + \int_{k_0}^\infty \zeta_{\mathrm{evan}}^\alpha k_\rho dk_\rho \right) \tag{B1}$$

where $k_0$ is the magnitude of the wavevector in the vacuum, $k_\rho$ is the parallel (to the interface) component of the wavevector, $\Theta$ is the mean energy of an electromagnetic state (47), TE (TM) stands for the transverse electric (transverse magnetic) polarization, and $\zeta_{\mathrm{prop}}^\alpha$ ($\zeta_{\mathrm{evan}}^\alpha$) is the transmission function for propagating (evanescent) waves with $\alpha$ polarization. The transmission functions $\zeta_{\mathrm{prop}}^\alpha$ and $\zeta_{\mathrm{evan}}^\alpha$ are calculated as (47):

$$\zeta_{\mathrm{prop}}^\alpha = \frac{\left(1 - |R_{0E}^\alpha|^2 - |T_{0E}^\alpha|^2\right)\left(1 - |R_{0R}^\alpha|^2 - |T_{0R}^\alpha|^2\right)}{\left|1 - R_{0E}^\alpha R_{0R}^\alpha e^{i2k_{0z}D}\right|^2} \tag{B2a}$$

$$\zeta_{\mathrm{evan}}^\alpha = \frac{4\mathrm{Im}(R_{0E}^\alpha)\mathrm{Im}(R_{0R}^\alpha)e^{-2\mathrm{Im}(k_{0z}D)}}{\left|1 - R_{0E}^\alpha R_{0R}^\alpha e^{i2k_{0z}D}\right|^2} \tag{B2b}$$

where $k_{0z}$ is the $z$-component of the wavevector in the vacuum, subscripts E and R refer to the emitting and receiving medium, respectively, and $R_{0j}^\alpha$ and $T_{0j}^\alpha$ are, respectively, the reflection and transmission coefficients of medium $j$ ($j$ = E or R) when illuminated by an incident electromagnetic field from the vacuum gap. $R_{0j}^\alpha$ and $T_{0j}^\alpha$ can be found using (47):

$$R_{0j}^{\alpha} = \frac{r_{0j}^{\alpha} + r_{js}^{\alpha} e^{2ik_{jz}t_j}}{1 + r_{0j}^{\alpha} r_{js}^{\alpha} e^{2ik_{jz}t_j}} \tag{B3a}$$

$$T_{0j}^{\alpha} = \frac{t_{0j}^{\alpha} t_{js}^{\alpha} e^{2ik_{jz}t_j}}{1 + r_{0j}^{\alpha} r_{js}^{\alpha} e^{2ik_{jz}t_j}} \tag{B3b}$$

where $k_{jz}$ denotes the $z$-component of the wavevector in medium $j$ ($j$ = E or R), $t_j$ is the thickness of medium $j$, subscript $s$ refers to the copper plates serving as the substrates for the emitter and receiver, and $r_{mn}^{\alpha}$ and $t_{mn}^{\alpha}$ are the Fresnel reflection and transmission coefficients at the interface of media $m$ and $n$ for $\alpha$ polarization, respectively, which are found as (47):

$$r_{mn}^{\text{TE}} = \frac{k_{mz} - k_{nz} - \omega \mu_0 \sigma_n}{k_{mz} + k_{nz} + \omega \mu_0 \sigma_n} \tag{B4a}$$

$$r_{mn}^{\text{TM}} = \frac{\varepsilon_{r,n} k_{mz} - \varepsilon_{r,m} k_{nz} + \sigma_n k_{mz} k_{nz}/\varepsilon_0 \omega}{\varepsilon_{r,n} k_{mz} + \varepsilon_{r,m} k_{nz} + \sigma_n k_{mz} k_{nz}/\varepsilon_0 \omega} \tag{B4b}$$

$$t_{mn}^{\text{TE}} = \frac{2k_{mz}}{k_{mz} + k_{nz} + \omega \mu_0 \sigma_n} \tag{B4c}$$

$$t_{mn}^{\text{TM}} = \sqrt{\frac{\varepsilon_{r,m}}{\varepsilon_{r,n}}} \frac{2\varepsilon_{r,n} k_{mz}}{\varepsilon_{r,n} k_{mz} + \varepsilon_{r,m} k_{nz} + \sigma k_{mz} k_{nz}/\varepsilon_0 \omega} \tag{B4d}$$

In Eqs. B4a-d, $\varepsilon_0$ and $\mu_0$ are the permittivity and permeability of the vacuum, respectively, $\varepsilon_{r,m}$ ($\varepsilon_{r,n}$) is the dielectric function of medium $m$ ($n$), and $\sigma_n$ is the electrical conductivity of graphene sheet covering medium $n$ ($\sigma_n$ = 0 if medium $n$ is not covered with a graphene sheet), which is found using the Kubo formula (47,48).

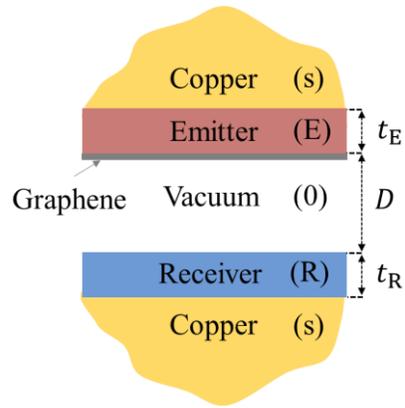

Figure 5 – The schematic considered for theoretical modeling of near-field radiative heat transfer for the system. A graphene-covered emitter (medium E) with a thickness of $t_E$ is separated from a receiver (medium R) with a thickness of $t_R$ by a vacuum gap (medium 0) of size $D$. The emitter and the receiver have temperatures $T_E$ and $T_R$, respectively, and are in contact with semi-infinite copper substrates (medium s).

### *Transferring Graphene to a LiF Plate*

A 60 mm by 40 mm monolayer of graphene CVD-grown on a copper foil and protected with a 60 nm layer of PMMA (Product#: ME0613, MSE supply) is transferred on a LiF plate using the standard wet transfer method (51,52) in a class 1000 cleanroom environment. For this purpose, the purchased graphene monolayer is cut into 15 mm by 15 mm pieces using a sharp razor blade. Then, the copper substrate is etched away by floating the sample on a 50% (by weight) solution of $FeCl_3$ in deionized (DI) water for approximately 15 minutes. Next, the PMMA-graphene film is washed in a DI water bath to remove the residual $FeCl_3$ solution from the sample. The washing process is repeated twice, after which the LiF plate is used to fish the PMMA-graphene film out of the bath. The sample is let to dry in the air for 2 hours and is then annealed in a hot chamber at a temperature of 150°C for 1 hour. The annealing process helps with removal of the water droplets and increasing the adhesion of graphene to the LiF substrate. To remove the PMMA coating, acetone is applied on the surface of the sample, and then the sample is rinsed with iso-propyl alcohol (IPA) to wash the residual acetone away. The sample is washed in the DI water and is let to dry in the air. The dried sample is then baked for 2 hours at 90℃ to remove the residual IPA and increase adhesion to the LiF plate.

## *Dispersion Relation of Coupled Surface Phonon and Plasmon Polaritons*

The dispersion relation of the coupled surface phonon and plasmon polaritons for a graphene covered dielectric medium (medium E) separated by a gap of size $D$ from a second dielectric medium (medium R) can be found as $e^{D\kappa_0}\left(\frac{\varepsilon_{r,R}}{\kappa_R} + \frac{1}{\kappa_0}\right)\left(\frac{\varepsilon_{r,E}}{\kappa_E} + i\frac{\sigma_E}{\omega\varepsilon_0} + \frac{1}{\kappa_0}\right) - e^{-D\kappa_0}\left(\frac{\varepsilon_{r,R}}{\kappa_R} - \frac{1}{\kappa_0}\right)\left(\frac{\varepsilon_{r,E}}{\kappa_E} + i\frac{\sigma_E}{\omega\varepsilon_0} - \frac{1}{\kappa_0}\right) = 0$ (53). In this equation, subscripts E and R refer to the emitting and receiving medium, respectively, and $\kappa_j = -ik_{jz}$.

## Associated Content

## *Supporting Information*

The Supporting Information is available free of charge at …

Optimal chemical potential for graphene, the maximal enhancement factor versus resonance misalignment, the effect of the number of graphene layers on the enhancement factor (PDF)

## Acknowledgment

The authors thank Stephen Abbadessa for the help with implementing the experimental setup.

## Funding Sources

This work is supported by the National Science Foundation under Grant No. CBET-2046630.

# SUPPORTING INFORMATION

## A. Optimal Chemical Potential for Graphene

The objective of this section is to study how the optimal chemical potential of graphene, $\mu_{c,\text{opt}}$, resulting in the largest enhancement of near-field heat flux, varies with the amount of resonance misalignment between the emitter and receiver. For this purpose, the optimal chemical potential for the heat flux between a LiF emitter and four different receivers (namely MgO, GaN, 6H-SiC, and SiO$_2$) with resonance frequency misalignments, $\Delta\omega$, ranging from $0.09\times10^{14}$ to $1.24\times10^{14}$ rad/s are calculated and shown in Fig. S1a. The temperatures of the emitter and receiver are assumed as 328.5 and 296.6 K, respectively, while a vacuum gap of size $D = 120$ nm is considered. The dielectric functions of the receivers are obtained from literature (S1-S3). It is seen from Fig. S1a that as $\Delta\omega$ increases, $\mu_{c,\text{opt}}$ increases. The reason can be explained by comparing the spectral heat flux per unit wavevector, $q_{\text{rad},\omega,k_\rho}$, for a receiver with a relatively small misalignment such as GaN with $\Delta\omega = 0.31\times10^{14}$ rad/s with that for a receiver with a larger misalignment such as SiO$_2$ with $\Delta\omega = 1.24\times10^{14}$ rad/s. Figures S1b and S1c show $q_{\text{rad},\omega,k_\rho}$ for TM-polarized electromagnetic waves in the absence of the graphene sheet for GaN and SiO$_2$ receivers, respectively, while Figs. S1d and S1e show the same for when LiF is covered with a graphene sheet with an optimal chemical potential (0.11 and 0.46 eV for GaN and SiO$_2$, respectively). It is seen from Figs. S1b and S1c that as $\Delta\omega$ increases, the wavevector of the surface modes with largest contribution to the heat flux at $\omega_{\text{SPhP}}$ of the receiver, $k_{\rho,\text{max}}$, decreases significantly ($k_{\rho,\text{max}}$ is marked in Figs. S1b and S1c for GaN and SiO$_2$ receivers, respectively). It is also observed from Figs. S1d and S1e that as the chemical potential of graphene increases, the slope of the upper branch of the coupled SPhP-SPPs associated with the graphene on LiF increases. To obtain the largest enhancement factor, the upper branch of the coupled SPhP-SPPs needs to intersect the dispersion branch associated with the SPhPs of the receiver around $k_{\rho,\max}$. As such and since $k_{\rho,\max}$ decreases with increasing $\Delta\omega$, a larger chemical potential is needed for $SiO_2$ with a larger misalignment of $\Delta\omega = 1.24\times10^{14}$ rad/s than GaN with $\Delta\omega = 0.31\times10^{14}$ rad/s. In conclusion, the larger the resonance misalignment, the greater the optimal chemical potential of graphene.

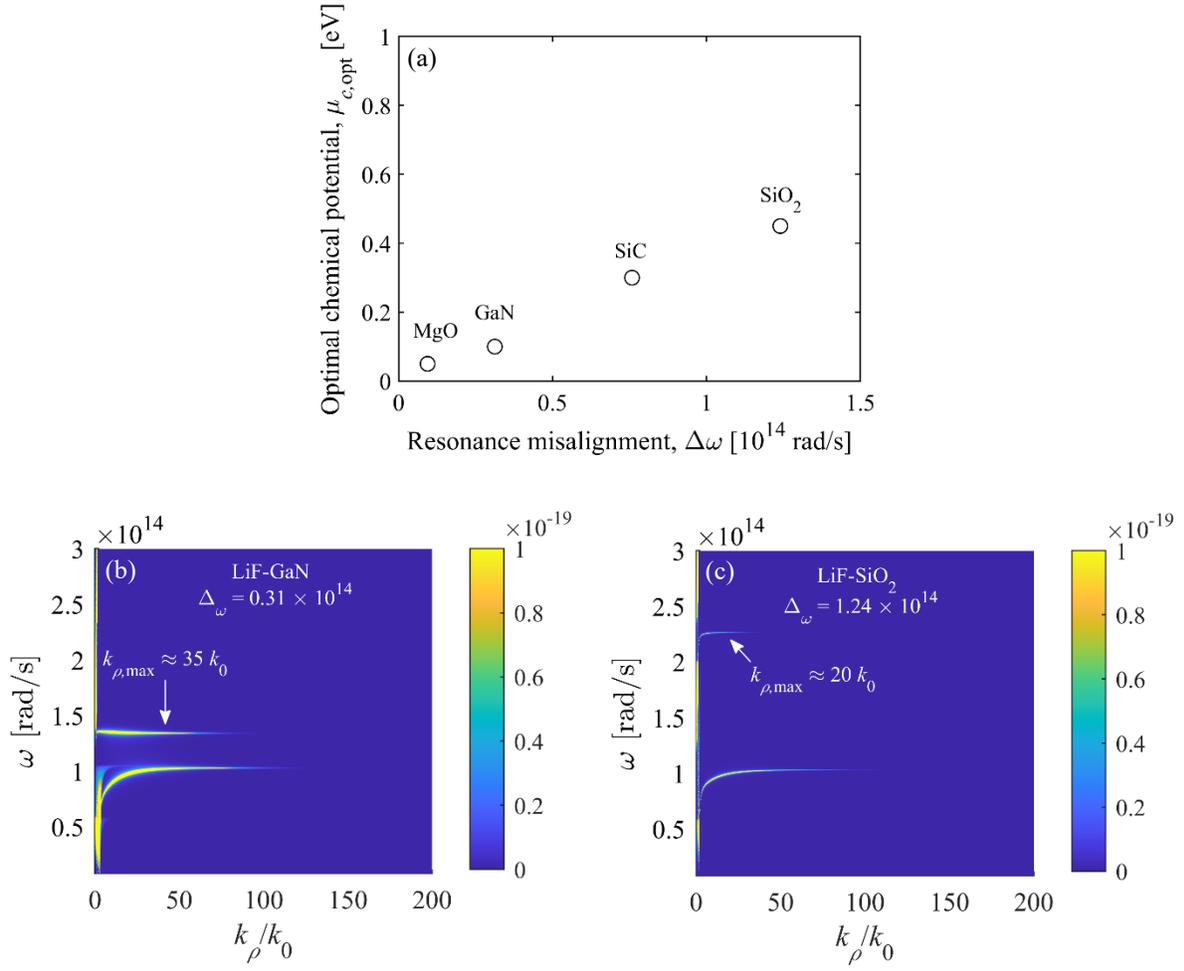

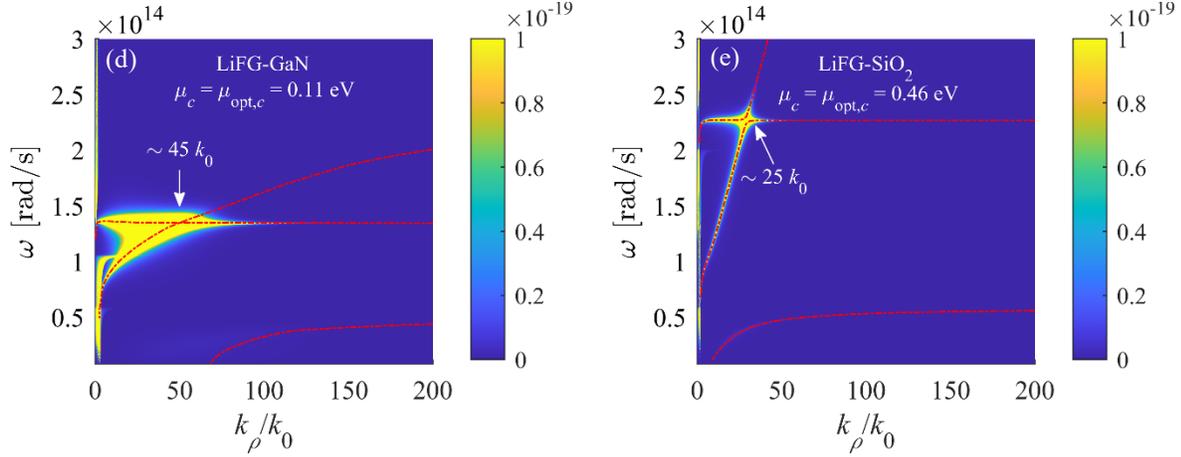

Figure S1 – (a) The optimal chemical potential, $\mu_{c,opt}$, for the heat flux between a graphene covered LiF (LiFG) substrate and a receiver with a resonance misalignment of $\Delta\omega$. The misalignment is measured relative to the SPhP frequency of LiF, which is equal to $1.03\times10^{14}$ rad/s. (b,c) The spectral heat flux mediated by the TM-polarized electromagnetic waves per unit $k_\rho$, $q_{rad,\omega,k_\rho}$, for a LiF emitter and (b) a GaN receiver with a resonance misalignment of $0.31\times10^{14}$ rad/s and (c) a SiO$_2$ receiver with a resonance misalignment of $1.24\times10^{14}$ rad/s. (d,e) The same as Panels (b,c) but for when LiF is covered with a graphene sheet with an optimal chemical potential. In Panels a-e, the temperatures of the emitter and receiver are assumed as 328.5 and 296.6 K, respectively, while the size of the vacuum gap is $D = 120$ nm. The unit for the color bars in Panel b-e is Wm$^{-2}$(rad/s)$^{-1}$m.

The strength of coupling between surface modes is strongly impacted by the chemical potential of graphene. The largest enhancement of heat flux is obtained for an optimal chemical potential which provides an ideal slope for the upper branch of the coupled SPhP-SPPs of LiFG to intersect the SPhP branch of the receiver around $k_{\rho,max}$. Figure S2a shows $q_{rad,\omega,k_\rho}$ mediated by the TM-polarized electromagnetic waves for the LiF-SiC system. It is seen from this figure that $k_{\rho,max} \approx$ $20k_0$. When $\mu_c = \mu_{c,\text{opt}}$ (see Fig. S2b), the upper branch of coupled SPhP-SPPs intersect the SPhP branch of SiC around $k_{\rho,\text{max}}$ resulting in a highly efficient surface modes. When $\mu_c < \mu_{c,\text{opt}}$ (see Fig. S2c and S2d), the slope of the upper branch is not sufficiently large, such that the upper branch either intersects the SPhP branch of the receiver at wavevectors greater than $k_{\rho,\text{max}}$ (Fig. S2c) or does not intersect the SPhP branch of the receiver at all (Fig. S2d). When $\mu_c > \mu_{c,\text{opt}}$, the slope of the upper branch increases resulting in an intersection at wavevectors smaller than $k_{\rho,\text{max}}$ (e.g., see Fig. S2e for the LiFG-SiC system at $\mu_c = 0.50$ eV).

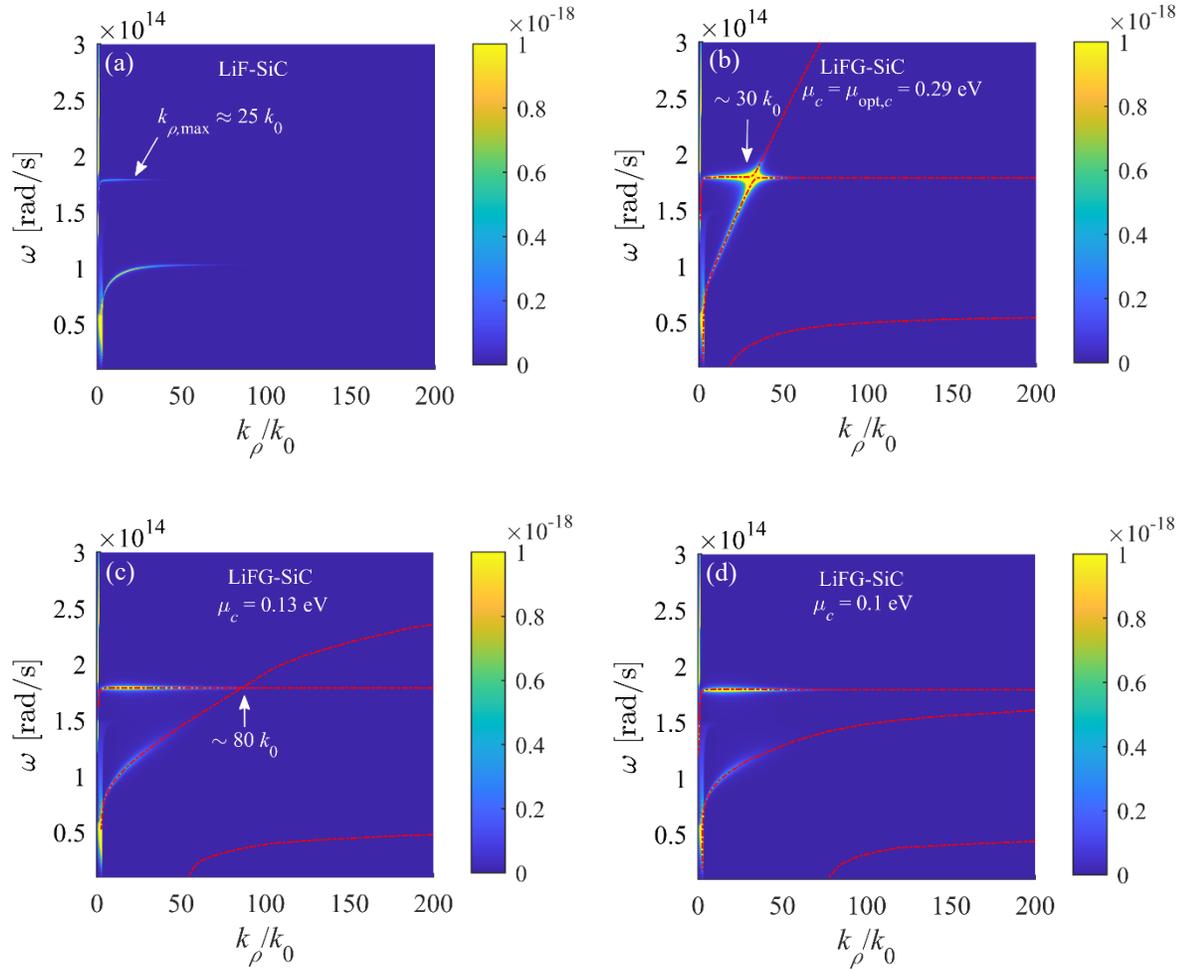

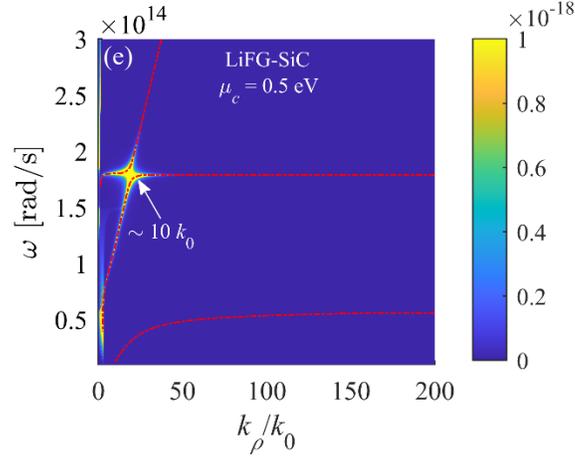

Figure S2 – The spectral heat flux mediated by the TM-polarized electromagnetic waves per unit $k_\rho$, $q_{\text{rad},\omega,k_\rho}$, for a LiF emitter at 328.5 K and a SiC receiver at 296.6 K separated by a vacuum gap of $D = 120$ nm. (a) No graphene. (b) $\mu_c = \mu_{c,\text{opt}} = 0.29$ eV (c) $\mu_c = 0.13$ eV. (d) $\mu_c = 0.10$ eV, and (e) $\mu_c = 0.50$ eV. The unit for the color bars in Panel a-e is Wm$^{-2}$(rad/s)$^{-1}$m.

## B. The Maximal Enhancement Factor versus Resonance Misalignment

The objective of this section is to analyze how the maximal enhancement factor obtained for heat flux between two dissimilar dielectric media is dependent on the amount of misalignment between resonance frequencies of the two materials. For this purpose, the maximal enhancement factor, $\eta_{\max}$, for the heat flux between LiF and eleven materials supporting SPhPs at different frequencies than LiF is calculated. The enhancement factor is defined as the ratio of near-field radiative heat flux in the presence of graphene to that in the absence of graphene. The temperatures of the emitter and receiver are 328.5 and 296.6 K, respectively, and a vacuum gap size of $D = 120$ nm is assumed. The dielectric functions of SiC and LiF are reported in the manuscript. For the other materials, the parameters of the Lorentz dielectric functions are taken from Refs. [47,S2] and a phonon scattering rate of $3.767 \times 10^{11}$ rad/s is considered. The enhancement factor for each material combination is dependent on the chemical potential of graphene. The maximal enhancement factor for each of the eleven cases versus the different between the SPhP frequencies of the emitter (LiF) and the receiver is shown in Fig. S3a. It is seen from Fig. S3a that the maximal enhancement factor for all cases is greater than 1 and ranges from 1.1 to 25.4, confirming the heat flux enhancement in the presence of graphene for all considered cases. The smallest enhancement factors pertain to the cases such as LiFG-InP where the SPhP resonance frequency of the receiver is located between the upper and lower branches of the coupled SPhP-SPPs of the graphene-covered emitter (LiFG), and thus there is no coupling between the surface modes across the vacuum gap. For example, the spectral heat flux per unit $k_\rho$, $q_{\text{rad},\omega,k_\rho}$, and the dispersion relation branches are plotted in Fig. S3b for an InP receiver for which $\eta_{\max} = 1.4$. The optimal chemical potential is assumed for graphene. In this case, the dispersion branch associated with the SPhPs of the receiver cannot intersect any of the branches associated with the coupled SPhP-SPPs of the LiFG, and thus the heat flux does increase appreciably. Small to moderate enhancement factors are obtained when there is only a slight misalignment between the resonance frequencies of the emitter and receiver. In these cases, the near-field heat flux is already large in the absence of graphene since the misalignment of resonance frequencies is small. The largest enhancement factors are obtained for the cases where there is a large misalignment between the SPhP resonances of the two materials. In these cases, the heat flux in the absence of graphene is quite low due the large misalignment of SPhP frequencies of the emitter and receiver. However, in the presence of the graphene sheet, the upper (Fig. S3c) or the lower (Fig. S3d) branch of the coupled SPhP-SPPs associated with LiFG crosses the SPhP dispersion branch of the receiver at large wavevectors, resulting in surface mode coupling across the vacuum gap. The coupling of surface modes across the vacuum gap resonantly enhances the heat flux at the $\omega_{\text{SPhP}}$ of the receiver.

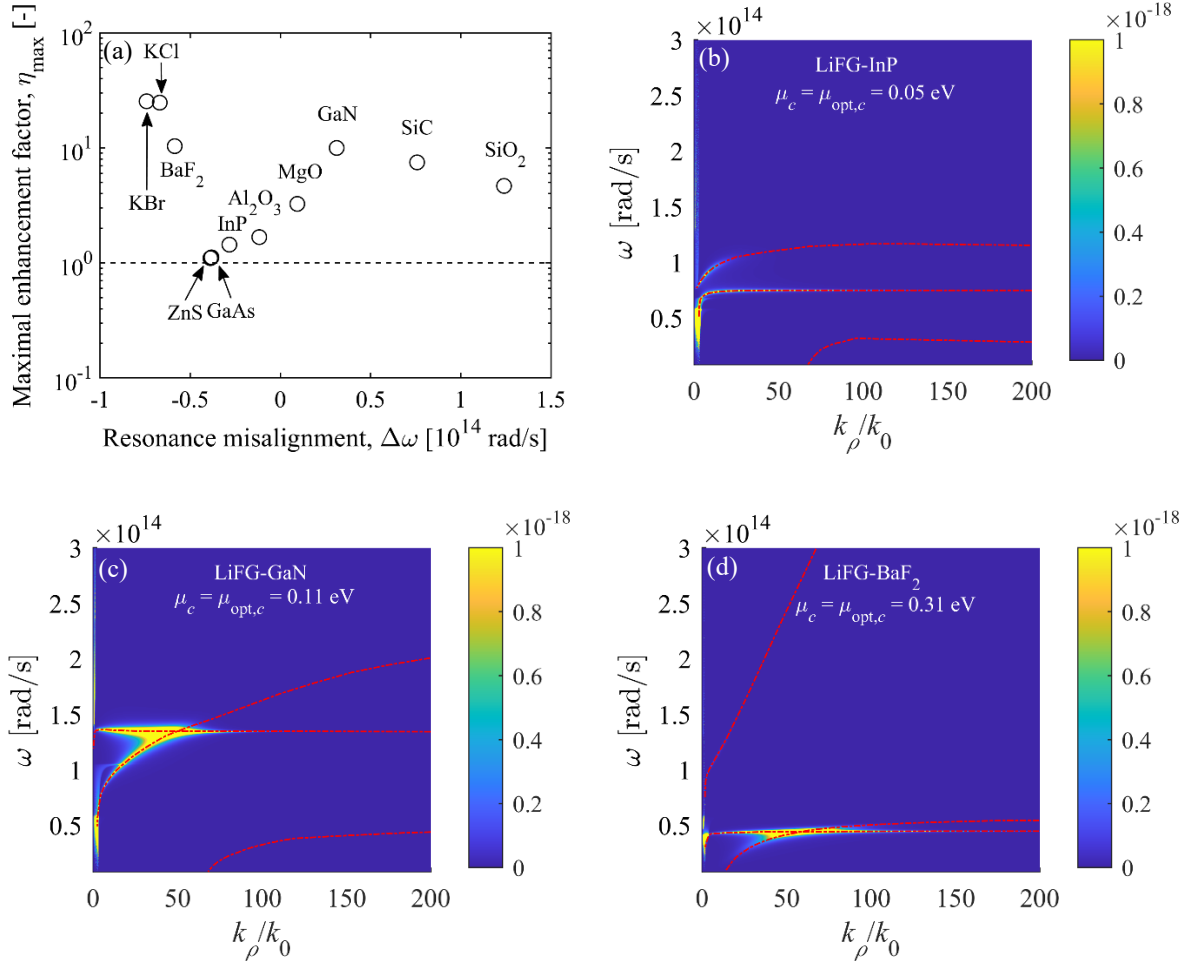

Figure S3 – (a) The maximal enhancement factor for radiative heat flux between graphene covered LiF (LiFG) and eleven materials supporting surface modes at various frequencies. (b,c,d) The spectral heat flux per unit wavevector, $q_{\text{rad},\omega,k_\rho}$, for LiFG-InP, LiFG-GaN, and LiFG-BaF$_2$ systems. The temperatures of the emitter and receiver are 328.5 and 296.6 K, respectively, and a vacuum gap size of $D$ = 120 nm is assumed. The unit for the color bars in Panel b-d is Wm$^{-2}$(rad/s)$^{-1}$m.

## C. The Effect of the Number of Graphene Layers on the Enhancement Factor

The objective of this section is to study the effect of number of graphene layers used on the emitter side on the enhancement factor of near-field heat flux. For this purpose, the maximal enhancement factor, $\eta_{\text{max}}$, for heat flux between a graphene-covered LiF substrate and a SiC substrate versus the number of graphene layers is calculated. The LiF substrate is covered with a graphene sheet while a vacuum gap of size 10 nm is assumed between subsequent graphene layers. The temperatures of the emitter and receiver are 328.5 and 296.6 K, respectively, and a vacuum gap size of $D = 120$ nm is assumed. The maximal enhancement factor and the optimal chemical potential, $\mu_{c\text{opt}}$, resulting in the maximal enhancement factor are shown in Fig. S4a. It is seen that the enhancement factor for the case with multilayers graphene is greater than the case where only one layer of graphene is used. The reason can be explained by considering Figs. S4b and S4c which compare the spectral heat flux per unit $k_\rho$, $q_{\text{rad},\omega,k_\rho}$, for the cases where one and three graphene layers are used. When three graphene sheets are used, the lowest dispersion branch associated with the coupled SPhP-SPPs of LiFG also acquires a positive slope. This positive slope enables the lower branch to reach and intersect the dispersion branch associated with the SPhPs of SiC, creating a region of highly efficient modes at large wavevectors enhancing the heat flux at the SPhP frequency of SiC substantially. Figure S4a also shows that increasing the number of graphene layers beyond seven does not change the enhancement factor. As the number of graphene layers increases, the effect of graphene layers adjacent to the LiF emitter on the heat transfer decreases and eventually vanishes.

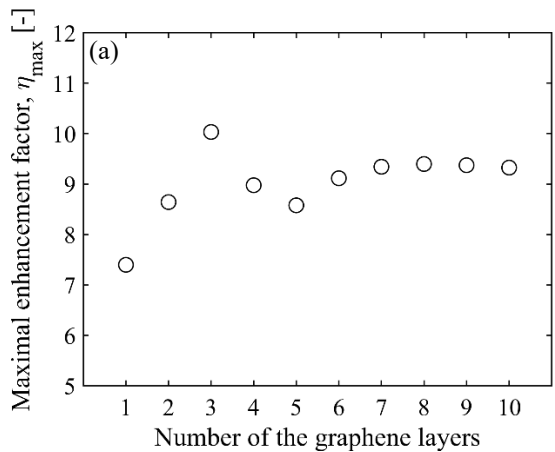

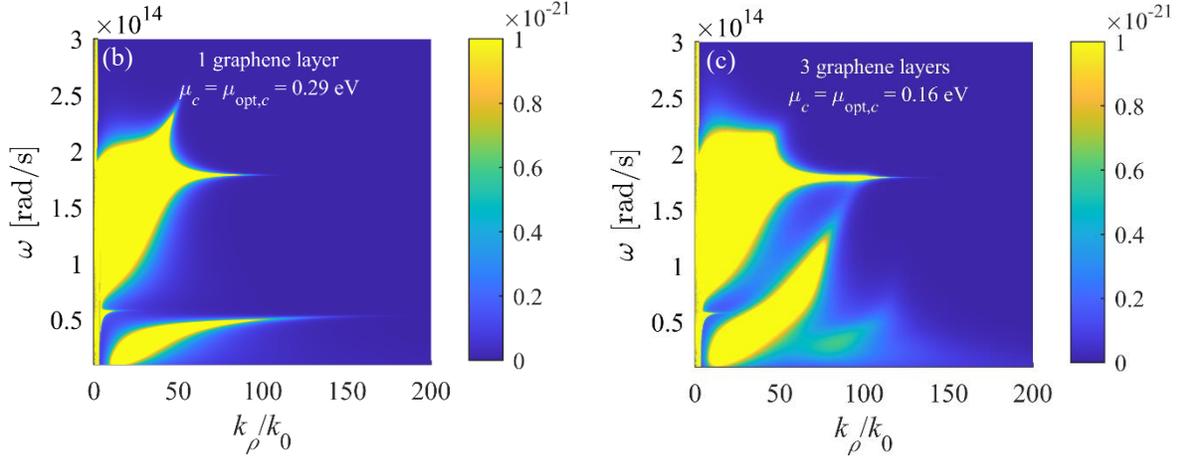

Figure S4 – (a) The maximal enhancement factor, $\eta_{max}$, for near-field radiative heat flux between graphene-covered LiF and SiC versus the number of graphene layers. The first graphene layer is in contact with the LiF substrate, while a vacuum gap of size 10 nm is assumed between the subsequent graphene layers. (b,c) The spectral heat flux per unit $k_\rho$, $q_{rad,\omega,k_\rho}$, mediated by the TM-polarized electromagnetic waves for the cases where (b) one and (c) three graphene layers are used. The unit for the color bars in Panels b and c is $Wm^{-2}(rad/s)^{-1}m$. In Panels a-c, the emitter and receiver are assumed at 328.5 and 296.6 K, respectively, and a vacuum gap of size $D$ = 120 nm is considered.